*Jet deflection by very weak guide fields during magnetic reconnection*

M. Goldman,[1] G. Lapenta,[2] D. Newman,[1] S. Markidis[2] and H. Che[1]

[1] *Dept. of Physics and CIPS, University of Colorado, Boulder, CO 80309, USA*

[2] *Centrum voor Plasma-Astrofysica, Katholicke Universiteit Leuven, Belgium*

**Abstract**

Simulations of anti-parallel reconnection have shown collimated electron jets outflowing from the x-point, and associated highly elongated "outer electron diffusion regions." New PIC simulations with ion/electron mass ratios as large as 1835 show that jets are deflected towards the magnetic separatrix by out-of-plane guide fields, $B_g$ as small as 0.05 times the asymptotic reconnecting field, $B_0$. The outer electron diffusion region is distorted and broken up, but the diffusion rate is unchanged. These results are interpreted in terms of electron dynamics and are compared to recent measurements of reconnection jets in the magnetosheath.

## Introduction

Magnetic reconnection is currently one of the most actively studied processes in plasma physics. It is responsible for solar flares, coronal mass ejections, magnetospheric substorms and sawtooth disruptions which limit plasma heating in toroidal fusion. The study of electron-scale processes associated with magnetic reconnection in the magnetosphere is one of NASA's highest priorities — to culminate in 2014 with the launch of the Multiscale Magnetosphere Satellites which will be able to resolve electron features of reconnection up to 100 times faster than existing satellites.

Among such features are electron flow velocity "jets," which are coherently directed, and highly extended spatially along the ion outflow exhausts (the "x"-direction). In recent simulations,[1,2,3] such electron jets, accompanied by elongated "outer diffusion regions" (regions where magnetic field lines are NOT frozen in to the plasma), were found during *antiparallel* reconnection — in which the initial reversing magnetic field, $\mathbf{B} = B_0 \tanh(y/\Delta y)\hat{x}$ lies entirely in the x-y plane.

It has been shown[4] that the addition of a small guide field, $B_g \hat{z}$ can drastically alter certain features of antiparallel reconnection. However, in this Letter, we provide the first evidence from PIC reconnection simulations[5] (with an ion to electron mass ratio of up to M/m = 1836) that guide fields of $B_g = 0.05 B_0$ and $0.1 B_0$ are sufficient to deflect and distort the jets from the x-axis towards separatrix legs, while the reconnection rate remains essentially unchanged. This effect is also present for M/m = 256 in a very large simulation box, in which it is seen to persist over long times – after the reconnection rate has reached a quasi-steady state. A simplified theoretical model of electron dynamics in a small region around the x-pt. shows how a small $B_g$ couples electron oscillations in y to growth in x, causing deflection of the electron jet.

## Reconnection simulations of jets in small guide fields

In order to determine how electron jets depend on the guide field, new implicit 2D PIC simulations[5] of spontaneous reconnection have been performed. The initial state is a perturbed[5] Harris equilibrium, with reversing component $B_x(y) = B_0 \cdot \tanh(y/\Delta y)$, where $\Delta y$ is the initial current sheet thickness, taken here to be $d_i/2$. Initial electron and ion temperatures are $T_{e0}/m_e c^2 = (v_{the0}/c)^2 = 2 \times 10^{-3}$, and $T_{i0} = 5T_{e0}$. The background density is $n_b = 0.1 n_0$, where $n_0$ is the peak initial density.

In Fig 1, electron flow velocities, $v_{ex}(x,y)/v_{the0}$, are compared at the same time, $\Omega_i t = 14.2$, in three reconnection simulations, for M/m = 1836. At this time the reconnection rate is declining from its peak value but has not yet reached a plateau (quasi-steady state). Later times will be considered later, for M/m = 256 (Fig. 4).

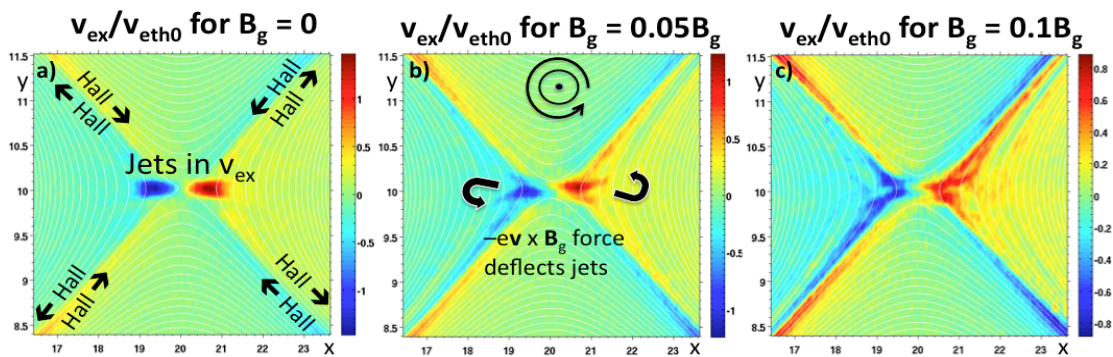

**Figure 1:** Deflection of jet in x-velocity of electrons, $v_{ex}/v_{the0}$, by small guide fields during reconnection (M/m = 1836). a) $B_g = 0$; b) $B_g/B_0 = 0.05$; c) $B_g/B_0 = 0.1$. x and y in units of $d_i$.

For $B_g$ =0 (Fig. 1a), incipient outgoing collimated jets flow out from the x-pt along the ±x-axis at y = $10d_i$. However, even for $B_g$ = $0.05B_0$, (Fig. 1b), the jets are split and deflected. For $B_g$ = $0.1B_g$ (Fig. 1c) the deflection of jets by the Lorentz force associated with $B_g$ has resulted in dominant beams elongated along appropriate separatrix legs.

Further out on the separatrix, $v_{ex}$ is part of the *electron Hall current* surrounding the *Hall B-field*. This velocity becomes more pronounced as the guide field is increased, consistent with the distortion of the Hall B-field pattern in x and y, recently reported for larger guide fields.[6]

The net electric plus magnetic force[1,2,7,8] on electrons at x and y is given by, **F**(x,y) = -e𝓔(x,y), where 𝓔 is defined as a generalized "field," 𝓔 = **E** + **v** x **B**/c. In Fig. 2, components of 𝓔(x,y) are plotted at the same time, $t\Omega_i$ = 14.2 as in Fig. 1.

Fig. 2a shows $𝓔_z$ for $B_g$ = 0. The in-flowing electron y-velocity, $v_{ey}$, is essentially frozen-in as ±$cE_z/B_x$ near x = $20d_i$ for y-values above ≈ $10.1d_i$ and below ≈ $9.9d_i$. The central (blue) region is the "inner" diffusion region and the flanking (red) regions are the early (not yet elongated) "outer" diffusion regions.[1,2]

The effect of a small guide field, $B_g$ = $.1B_0$ on $𝓔_z$ is shown in Fig. 2b. The inner diffusion region is distorted, but more significantly, the outer diffusion region is no longer elongated in x or collimated in y. It has been fragmented, tilted and distorted. However, the reconnection rate (proportional to the reconnection electric field, |$E_z$,|) is

essentially *unchanged*. We show in a later section that the reconnection *rate* is also unaffected at later times, even though jets in $v_{ex}$ remain distorted and deflected.

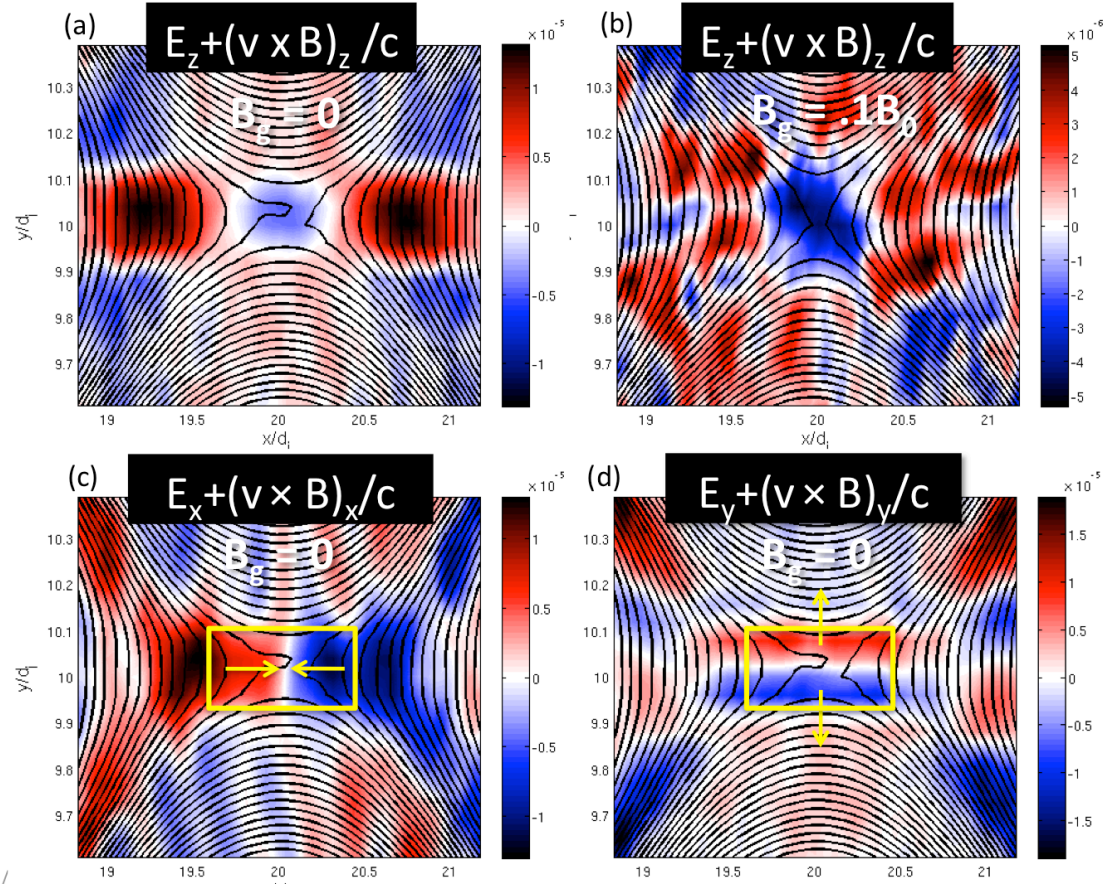

**Figure 2:** Components of electric-magnetic force "field," $\mathcal{E}(x, y) = \mathbf{F}/(-e) = \mathbf{E}+(\mathbf{v} \times \mathbf{B})/c$ in simulation units, $\sqrt{4\pi n_0 m_i c^2}$ (M/m = 1836, $t\Omega_i$ = 14.2). a) $\mathcal{E}_z$ for $B_g$ = 0; b) Same for $B_g$ = 0.1$B_0$, c) $\mathcal{E}_x$ and d) $\mathcal{E}_y$, for $B_g$ =0. Yellow arrows show direction of $\mathcal{E}_x$ and $\mathcal{E}_y$. Rectangles enclose "linear" regions.

Figs. 2c and 2d show $\mathcal{E}_x$ and $\mathcal{E}_y$, respectively, for $B_g$ = 0. The yellow rectangles enclose a region in which $\mathcal{E}_x$ is approximately linear and odd in x and $\mathcal{E}_y$ is approximately linear and odd in y. The boundaries of the rectangles at y = ±0.09$d_i$ are essentially the y-boundaries of the jets in Fig. 1a which are also the y-boundaries of the *current sheet*. The boundaries at x = ±0.4$d_i$ are at about half the values of of the x-boundaries of the jets. In this region the force, $-e\mathcal{E}_x \propto x$, accelerates electrons *away from the x-pt.* at x =

$20d_i$ to form the outgoing jets. The Hall field, $B_z$ is negligible here and the dominant contribution to $-e\mathcal{E}_x$ is the Lorentz force, $ev_{ez}B_y(x)$, produced by the out-of-plane almost-uniform velocity $v_{ez}$ of the electron *current sheet* and the in-plane $B_y(x)$ of the *reconnected* field lines.[1,2]

The *electric force,* $-eE_x$, opposes $+ev_{ez}B_y/c$ in Fig. 2c, and cancels about half of it, leaving a net outward-accelerating force at this early time. $E_x$ is the so-called *electron trapping field*,[2,7] which dominates in regions outside of the yellow rectangle and leads to oscillations in x in both the in-flow region and the exhaust.[7] In the linear region in Fig. 2d, $-e\mathcal{E}_y(y) \propto -y$ is a *restoring* force in y, which *traps the jet in the y-direction* and keeps it from expanding in y (for antiparallel reconnection). Here, again, the electric and magnetic parts of $\mathcal{E}_y$ oppose each other and, once again, the Lorentz force, $ev_{ez}B_x(y)/c$ is about twice as big as the *(Hall) electric field* force, $-eE_y$.

## Electron dynamics in linear region

The dynamical equations for 2-D in-plane motion of an electron due to the in-plane fields $E_{x,y}(x,y)$ and $B_{x,y}(x,y)$ underlying Fig. 2c and 2d, together with gyromotion at $\Omega_{eg} = eB_g/m_ec$ in a guide field, $B_g \leq 0.1B_0$ are:

(1a) $\ddot{x} + \dfrac{e}{m}\left[E_x(x,y) - \dfrac{\dot{z}B_y(x,y)}{c}\right] = -\Omega_{eg}\dot{y},$

(1b) $\ddot{y} + \dfrac{e}{m}\left[E_x(x,y) + \dfrac{\dot{z}B_x(x,y)}{c}\right] = \Omega_{eg}\dot{x}$

Assume now that $\dot{z}$ is equal to its initial value, $\dot{z}_0$, which, in turn, is set equal to $v_{ez}$ of the current sheet in the linear region. where it is essentially independent of x and y. Neglect of the dynamical evolution of $\dot{z}$ away from its initial value, $\dot{z}_0$ sets this treatment apart from that of the so-called Speiser orbits.[9] This assumption has been verified by tracking the exact dynamics of $\dot{z}$, starting from $\dot{z}_0$ in the linear region. The square brackets in eqns. (1) are now become, respectively, $\mathcal{E}_x(x,y)$ and $\mathcal{E}_y(x,y)$ in the linear region in Fig. 2c,d. These in-plane forces for $B_g = 0$ drive electron motion to an adequate approximation even in the presence of sufficiently small guide fields. It will be verified later that $\mathcal{E}_x$ and $\mathcal{E}_y$ are approximately stationary and equal to their values at time 14.2 throughout the electron motion in the linear region. We now make explicit *linear*[1] approximations for both $\mathcal{E}_x(x)$ and $\mathcal{E}_y(y)$ in this region, yielding a representation of the forces that is adequate to capture the physics of jet deflection due to coupling of x and y motions in the linear region by electron gyromotion, $\Omega_{eg}$:

(2) $\ddot{x} - \gamma_x^2 x \approx -\Omega_{eg}\dot{y}, \quad \ddot{y} + \omega_y^2 y \approx \Omega_{eg}\dot{x},$

The coefficients $\gamma_x^2 \equiv -\omega_e^2 \left[\partial_{x'} \mathcal{E}_x^{sim}(x')\right]_{x'=0} > 0$ and $\omega_y^2 \equiv \omega_e^2 \left[\partial_{y'} \mathcal{E}_y^{sim}(y')\right]_{y'=0} > 0$. The origin of the spatial variables x and y has been relocated to the x-point. Distances are in ion inertial lengths, $d_i$. From Figs. 2c,d, the rates are roughly, $\gamma_x/\Omega_{e0} = 0.035$ and $\omega_y/\Omega_{e0} = 0.066$, where $\Omega_{e0}$ is the electron cyclotron frequency in $B_0$. The initial conditions are chosen to be $x_0 = 0$ and $y_0 = 0.09 d_i$ (close to the top of the rectangle of validity). Consistent with the frozen-in electron E x B drift velocity, the initial y-velocity at $y_0$ is taken from the simulation to be $\dot{y}_0 = -0.2 v_e$. At this point a downward moving electron begins to feel the y-force, $(-e/m)\mathcal{E}_y(y)$, which eventually turns it back up again in y and keeps it confined in y as it is accelerated in x within the jet. A small initial velocity, $\dot{x}_0 = 0.2 v_{e0}$ is chosen, consistent with trapping[7] in x as an electron drifts down in y.

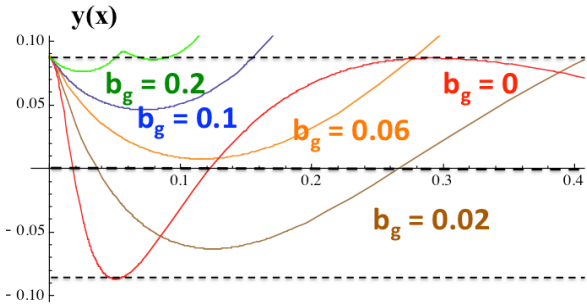

**Figure 3** Parametric plots of y(x) sol'ns to Eqns (2) for $b_g = B_g/B_0 = 0, 0.02, 0.06, 0.1, 0.2$. M/m = 1836

When $B_g = 0$, the solution to Eqns. 1 is $x(t) \propto \sinh(\gamma_x t)$ and $y(t) \propto \cos(\omega_y t)$. For $B_g = 0.1 B_0$, both x and y are linear combinations of $\sinh(\gamma' t)$, $\cosh(\gamma' t)$, $\sin(\omega' t)$ and $\cos(\omega' t)$, so that y is now growing as well as oscillating. Here, $\omega'$ and $\gamma'$ arise from the solutions of the eigenvalue equation for the linear eqns. (2):

$2\{\omega'^2, -\gamma'^2\} = \left[\Omega_{eg}^2 + (\omega_y^2 - \gamma_x^2)\right] \pm \sqrt{4\omega_y^2 \gamma_x^2 + \left[\Omega_{eg}^2 + (\omega_y^2 - \gamma_x^2)\right]^2}$. In Fig. 3 the *parametric trajectories* y(x) are plotted for $B_g/B_0 = 0, 0.02, 0.06, 0.1$ and $0.2$. For $B_g = 0$, the trajectory is bounded in y and extends out in x. For $B_g \neq 0$ the trajectories are all unbounded in y and leave the jet region at smaller and smaller x as $B_g$ increases. In order to justify the use of the *time-independent* forces $\mathcal{E}_x$ and $\mathcal{E}_y$, note that if t = 0 at x = 0 the crossing for $B_g$ = 0.1$B_0$ occurs at $t_c\Omega_i$ = 0.027.

## Reconnection simulations of jets in a guide field at later times

In order to study jets and diffusion regions at *later* times we have performed additional implicit PIC reconnection simulations with a mass ratio of M/m = 256 in a much larger simulation box of size 30$d_i$ x 200$d_i$. Time-histories of the reconnection rate ($\propto -E_z(t)$) at the x-pt. are shown in Fig. 4, together with snapshots of $v_{ex}/v_{eth0}$.

For $B_g$ = 0 (Figs. 4b,c,d) the jets are well-collimated and undeflected for as long as the simulation is run ($t\Omega_i$ = 48). Just before $t\Omega_i$ = 35 a secondary island[2] is formed to the right of the x-point which produces large oscillations in -$E_z(t)$.

With a guide field of $B_g$ = 0.1$B_0$ the jet behavior is quite different. At time 14.1 the jet deflection (Fig. 4e) looks quite similar to that for M/m = 1836 (Fig. 1c), suggesting there is no strong mass dependence for early jet deflection in a small guide field. At $t\Omega_i$ = 28.1, the plot (Fig. 4f) of $v_{ex}$ in the presence of $B_g$ = 0.1$B_0$ differs significantly from the plot (Fig. 4c) at the corresponding time for $B_g$ = 0. The jets in $v_{ex}$ near the x-pt. in Fig. 4f are almost completely suppressed, with only two short weak deflected branches present. This is a significant time because plateau formation has set in for -$E_z(t)$ by this time for the guide-field reconnection . Finally, at time 35.4, the jets in $v_{ex}$ in guide-field reconnection (Fig. 4g) are deflected and longer than at 28.1 but not more intense (note change in color table scale). It is clear from Fig. 4a that even though the jets are strongly distorted the reconnection rate is essentially the same at late times with or without the guide field.

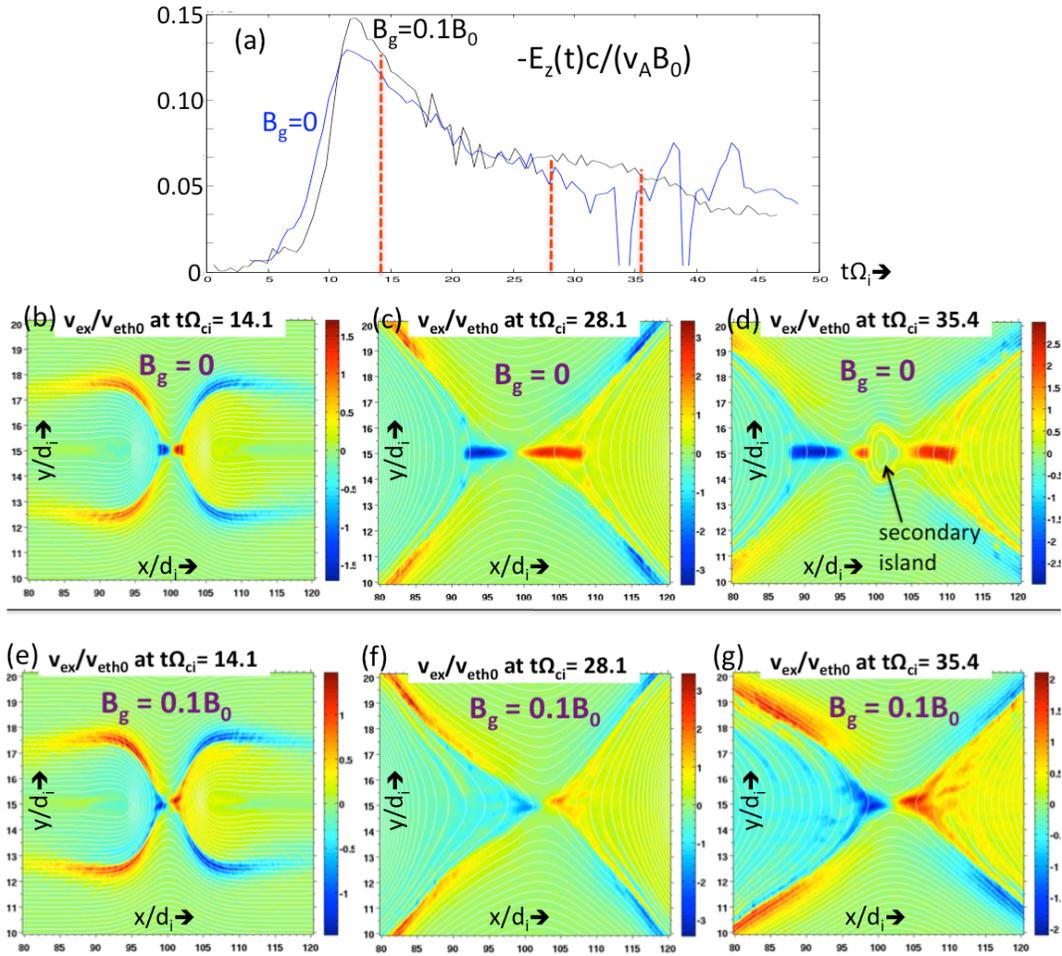

**Figure 4:** M/m = 256. (a): Reconnection rate, $-E_z(t)$, in units of inflow drift over inflow Alfven speed at x-pt. for $B_g = 0$ (blue); $B_g = 0.1B_0$ (black). Dashed vertical red lines show $t\Omega_i$= 14.1, 28.1 and 35.4 for which $v_{ex}(x,y)/v_{eth0}$ is plotted for $B_g = 0$ in (b), (c), (d) and for $B_g = .1$ in (e), (f), (d)

## Summary and significance

New implicit PIC simulations of reconnection have revealed that unexpectedly small guide fields can distort and deflect collimated electron jets in $v_{ex}$ away from the x-axis. For a physical mass ratio of M/m = 1836, a guide field, $B_g$, as small as one-twentieth of the asymptotic reversing field in the initial Harris equilibirum, $B_0$, is sufficient to deflect early-forming jets. A dynamical treatment of electron dynamics very close to the x-point during small guide field reconnection in a hydrogen plasma reveals the mechanism by which the Lorentz force in the guide field produces jet deflection. For M/m = 256 early jet disruption by a guide field $B_g = 0.1 B_0$ is sustained over long times, after the reconnection rate has begun to flatten. The time-history of the reconnection rate remains unchanged with the addition of such a small guide field, even though the narrow well-collimated "outer diffusion region" of antiparallel reconnection is destroyed along with the jets.

The significance of these new results is at least two-fold:  1) guide-field jet deflection is at odds with certain recent measurements[10] of undeflected electron x-jets in the presence of measured guide fields, and (2) this discrepancy may suggest that  initial conditions used in almost all spontaneous reconnection simulations be re-examined. Cluster measurements[10] reported in PRL, of an elongated outer diffusion region during symmetric magnetosheath reconnection, have revealed highly elongated ±x-directed electron jets  and external diffusion regions similar to those found in explicit

PIC simulations[1,2] with *no* guide field. The new reconnection simulations in this Letter should be *better* suited for comparison with the Cluster measurements since they use ion to electron mass ratio of up to 1836 (as opposed to M/m = 25 in Ref. 1) and a guide field, $B_g$, even *smaller than the measured* $B_g = 0.16B_0$. Surprisingly, however, the new simulations show that electron jets do *not* extend many ion inertial lengths in the x-direction, but rather are deflected towards the magnetic separatrix, as seen in earlier magnetosphere measurements.[11]

How can this difference be reconciled? The 2D simulations reported here (as well as the 2D simulations in Refs. 1 and 2) use an initial electron temperature, $T_e$ of 1keV. Although $T_e$ is not measured by Cluster, a best guess is $T_e \approx 40$eV (T. Phan, private communication). Reconnection at such cold temperatures are too computationally intensive to be simulated for realistic M/m, but implicit simulations at 250 eV for M/m = 256 and $B_g = 0.1B_0$ not discussed here continue to show jet deflection. At any rate, it seems unlikely that still colder temperatures would inhibit jet deflection because a lower temperature yields a tighter electron gyroradius in $B_g$, which suggests stronger, not weaker jet deflection. It also seems unlikely that 3D reconnection simulations will favor collimated electron jets in x, although this cannot be ruled out. More fundamental causes of the discrepancy include either special (as yet unknown) prevailing features of the magnetosheath at the time of measurement which favored undeflected long jets or limitations of the commonly used Harris equilibrium for intiating reconnection simulations (e.g., will *driven* reconnection show more robust long jets?). Whatever the resolution, almost all real reconnection events involve small guide fields and their effect on electron jets must be understood.

Understanding electron features of reconnection will be one of the main thrusts of the upcoming NASA Magnetosheric Multiscale (MMS) mission.

## Acknowledgement

This work was supported by NASA MMS Grant NNX08AO84G. We wish to thank Jan Egedal, Tai Phan, Forest Mozer and Michael Shay for useful discussions.